\documentclass[psfig]{kluwer}    

\usepackage[]{graphicx}

\newcommand{\etal}      {et al.\ }
\newcommand{\Ko}        {K\"or\"osmezey}
\newcommand{\BoMo}      {Bockel\'ee-Morvan}

\newdisplay{guess}{Conjecture}

\begin{document}                                                        
\begin{article}
\begin{opening}         
\title{Chemical Processes in Cometary Comae} 

\author{S.B. \surname{Charnley}$^{1}$, S.D. \surname{Rodgers}$^{1}$,
H.M. \surname{Butner}$^{2}$ and P. \surname{Ehrenfreund}$^{3}$}

\runningauthor{Charnley et al.}
\runningtitle{Coma Chemistry}

\institute{$^{1}$ Space Science Division, NASA Ames Research Center,  
MS 245-3, Moffett Field, CA 94035, USA \\
$^{2}$ SMTO-Steward Observatory, University of Arizona,
933 N. Cherry Avenue, Tucson, AZ 85721, USA \\
$^{3}$ Leiden Observatory, PO Box 9513,  2300 RA  Leiden, The Netherlands
}

\date{April 15 2002}

\begin{abstract}
Recent developments in the chemical modelling of cometary comae are
described. We discuss the cyanide chemistry and present new HCN
observations of the recent comet C/2002 C1 (Ikeya-Zhang).  The
connection between interstellar and cometary organic molecules is
discussed from the perspective of recent theories of interstellar
gas-grain chemistry.
\end{abstract}

\end{opening}

\section{Introduction}

Comets are the most volatile-rich and pristine objects in our solar
system. Although comets will have undergone some processing during
their $\sim$ 4.5~Gyr lifetime, their chemical composition will
fundamentally reflect the conditions in the outer solar nebula when
they were formed (Mumma \etal 1993, Irvine \etal 2000).  Determining
molecular abundances in cometary ices therefore allows us to place
important constraints on models of the protosolar nebula, and
planetary formation in general.  A key question, when considering
exactly when, where and how comets formed, is to what extent comets
consist of relatively pure interstellar material?  Detailed studies of
interstellar solid state absorption bands made with the {\it Infrared
Space Observatory (ISO)}, theoretical and observational progress in
the understanding of interstellar gas-grain chemistry, and the recent
apparitions of bright comets, particularly Hale-Bopp, have all
contributed to a deeper understanding of the chemical similarities and
differences between interstellar and cometary material (Table 1;
Ehrenfreund \& Charnley 2000).  Until the {\it Rosetta} and {\it Deep
Impact } space missions, we have no direct way to determine the
composition of the nucleus and must rely on gas phase observations of
coma molecules to estimate nuclear molecular abundances. However, a
great deal of chemical processing can occur in the coma and so
accurate chemical models are of importance when using coma observations
to derive nuclear ice composition.

\begin{table}
\caption{ Representative compositions of the gas in a cold molecular
cloud (L134N), protostellar ices (NGC7538:IRS9), protostellar hot core
gas (Sgr B2(N)), and in a cometary coma (Hale-Bopp).}
\begin{center}\scriptsize
\begin{tabular}{lllll}
\noalign{\smallskip}
\hline 
\noalign{\smallskip}
Molecule& L134N & NGC7538:IRS9 & Sgr B2(N) & Hale-Bopp \\
\noalign{\smallskip}
\hline
\noalign{\smallskip}
H$_2$O &$<3$  & 100   & $>$ 100  & 100\\
CO     &1000 & 16  &  1000 &  20 \\
CO$_2$ & - & 20 & -  & 6-20\\
H$_2$CO &0.25 & 5 & $>0.005$  & 1\\
CH$_3$OH &0.04 & 5  & 2 & 2\\
NH$_3$ & 2.5 &13 & - & 0.7-1.8\\
CH$_4$ &- & 2 & - & 0.6\\
C$_2$H$_2$ & - & $<$10 & - & 0.1\\
C$_2$H$_6$ & -&  $<$ 0.4 & - & 0.3\\
HCOOH     &0.004 &  3 & $>0.003$ & 0.06\\
CH$_2$CO& $<$0.009& - &0.002& $<0.03$\\
CH$_3$CHO& 0.008 & - &0.002 & 0.02\\
$c$-C$_2$H$_4$O &-& - &0.001 &-\\
CH$_3$CH$_2$OH&- &$<$1.2 &0.01 & $<0.05$\\
CH$_3$OCH$_3$&-& - &0.03 & $<0.45$\\
HCOOCH$_3$ & $<$0.02 & - &0.02 & 0.06\\
CH$_3$COOH & - & - & 0.0008& - \\
CH$_2$OHCHO & - & - & 0.003 & - \\
OCN$^-$ & - & 1 & - & -\\
HNCO & - & - & 0.006 & 0.06-0.1\\
NH$_2$CHO  & $<$0.001 & - & 0.002 & 0.01\\
HCN   & 0.05 &- & $>0.05$  & 0.25\\
HNC &0.08 &-& $>0.001$  &0.04\\
CH$_3$CN &$<$0.01  & - & 0.3 & 0.02\\
CH$_3$NC & - & - & 0.015 & - \\
HC$_3$N  &0.002  & - &0.05 & 0.02\\
C$_2$H$_3$CN & - & - & 0.6 & - \\
C$_2$H$_5$CN & - & - & 0.006 & - \\
H$_2$S  & 0.01 & - & - & 1.5\\
OCS & 0.02 &0.05 & $>$0.02 & 0.5\\
H$_2$CS & 0.008 & - &0.2 & 0.02 \\
SO& 0.25 & - & 0.2 & 0.2-0.8\\
SO$_2$ & 0.005 & - & 0.3 & 0.1\\
\noalign{\smallskip} \hline \noalign{\smallskip}
\end{tabular} \\
\end{center}
{\scriptsize NOTE: Abundances for interstellar ices and volatiles in
comet Hale-Bopp are normalized to H$_2$O. Gaseous abundances for the
Sgr B2(N) hot core and L134N are normalized to CO. Adapted from
Ehrenfreund \& Charnley (2000) and Charnley et al. (2002).}
\end{table}


\section{Modelling Coma Chemistry}

Following sublimation from the nucleus, parent molecules are exposed
to the solar UV field and they are photodissociated and ionized, creating
distinct plasma and neutral fluid components, as well as populations
of supra-thermal hydrogen atoms and molecules.  The initial expansion
of the gas is adiabatic and the temperature falls. Further out,
photodissociation and ionization reactions heat the gas and the
temperatures rise. A rich chemistry can ensue in the coma, involving
processes such as electron impact ionization/dissociation, charge
transfer, proton transfer, ion-atom interchange, and neutral
rearrangement (see Schmidt \etal 1988). Also, sublimation from and/or
photodestruction of dust injects new molecules into the coma.

Hydrodynamics and chemistry are intimately linked. Figure 1 shows that
many processes affect the thermal balance of the neutral
fluid. Chemistry can affect hydrodynamics: exothermic ($\sim1$~eV)
proton transfer reactions are an important heat source in the inner
coma, and cause the ion temperature to decouple from the neutral and
electron temperatures (\Ko\ \etal 1987).  Hydrodynamics can affect the
chemistry: large variations in the electron temperature throughout
the coma strongly affect the local rates of ion-electron recombination
reactions.

\begin{figure*}
\centerline{\includegraphics[width=285pt]{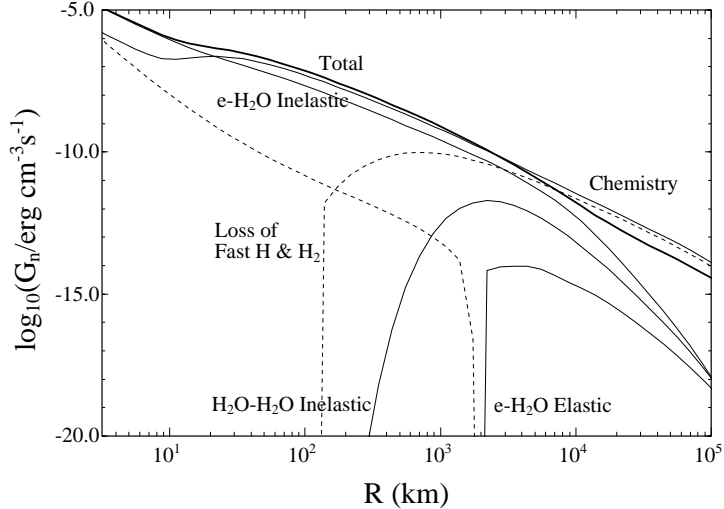}}
\caption{Major heating (full lines) and cooling (broken lines)
mechanisms for coma neutrals (from Rodgers \& Charnley 2002a)}
\end{figure*}

We have developed a hydrodynamical-chemical model of the coma (Rodgers
\& Charnley 2002a). A multifluid treatment is necessary and the model
calculates the coupled chemical and dynamical evolution in the
outflowing coma, computing in a self-consistent fashion molecular
abundances, individual densities and temperatures for the plasma
components and neutrals, as well as the inclusion of fast hydrogen atoms
as a distinct component. It is similar to earlier models which also
employed multifluid hydrodynamics (e.g.\ Huebner 1985; K\"or\"osmezey
\etal 1987) but, more pragmatically, we have the capability to study a
wide range of different chemistries with relative ease. This model has
been the basis of several recent chemical studies (Rodgers \& Charnley
1998, 2001b,c).


\section{HCN and HNC in Comets}

The HNC/HCN ratio varies greatly in different regions of the
interstellar medium, with an inverse temperature dependence: in cold
dark clouds the ratio can be higher than unity, whereas in hot cores
it is small.  At present we are unable to explain the excess of HNC in
cold clouds (see Talbi \etal 1996).  HCN has been seen in many comets,
but HNC was detected for the first time in comet Hyakutake, where it
was present with an abundance relative to HCN of 6\% (Irvine \etal
1996). HNC was subsequently detected in comets Hale-Bopp and Lee, and
the HNC/HCN ratio in Hale-Bopp showed a strong increase as the comet
approached perihelion, from $<2\%$ at 2.5~AU to $\sim 16\%$ at 1~AU
(Irvine \etal 1998). The Hyakutake results were originally interpreted
as proof that cometary ices contain unprocessed material from the ISM,
but the observations of Hale-Bopp proved conclusively that HNC must be
a daughter species, and that its production is related to the solar
photon flux and/or coma temperature.

Theoretical models can provide insight into possible chemical
production routes for HNC in the coma (Rodgers \& Charnley 1998,
2001c).  One must be careful to consider proton transfer reactions and
to add or exclude all exoergic or endoergic reactions of this type
(Hunter \& Lias 1998).  This has the important consequence that
ion-molecule chemistry is unable to synthesise the observed quantities
of HNC since, for example, HCNH$^+$ will react with methanol to form
HCN, but not HNC, thus quenching HNC production in methanol-rich
comets (Rodgers \& Charnley 1998).  However, endoergic isomerisation
reactions of HCN, driven by suprathermal H atoms produced in the
photodissociation of parent molecules, may be efficient in large,
active comets such as Hale-Bopp.

The observed HNC/HCN ratio in Comet Lee, of about 12\%, effectively
excludes bimolecular gas-phase reactions as the source of HNC in
comets less active than Hale-Bopp.  Detailed theoretical modelling
indicates that the HNC observed in comets Hyakutake and Lee must be
produced via photodissociation of some unknown parent(s), either
organic dust particles or large molecules (Rodgers \& Charnley 2001c).
Most small molecular candidates can be ruled out (e.g. HNCO,
CH$_2$NH).  Photo-fragmentation of large organic molecules/particles
has been proposed as an explanation of extended coma distributions of
several molecules and radicals, e.g. CO, C$_2$, C$_3$, H$_2$CO, CN and
NH$_2$\@.  It appears likely that this process is also a common source
of HNC, as well as perhaps accounting for the extended sources of HCN
and CN. Potential candidates for the unknown parent(s) include
hexamethylenetetramine (Bernstein \etal 1995) and
polyaminocyanomethylene (Rettig \etal 1992).

These analyses suggest there may be two viable sources of HNC in
comets. Photo-degradation being the most common, but with
isomerisation reactions driven by fast H atoms playing an important
role in large, active comets.  Clearly, one would wish to know the
HNC/HCN ratio, and its heliocentric variation, in many more
comets. Mapping the spatial distribution of HCN and HNC (e.g. Veal et
al.\ 2000; Blake et al.\ 1999) may also shed light on their origin and
relationship with other coma molecules.

\subsection{HCN in Comet Ikeya-Zhang }

Using the Kitt Peak 12m telescope and the Submillimeter Telescope
Observatory (SMTO) we have mapped the HCN emission from the coma of
the recent bright Comet C/2002 C1 (Ikeya-Zhang).  The comet ephemerides
used were based on IAU Circulars and were calculated and updated as
new circulars were released.  The millimeter observations at the 12m
had typical system temperatures of less than 800 K for 1 mm
observations.  The backends used were 100 kHz filterbanks and 250 kHz
filterbanks in parallel configuration.  Calibration was done with a
vane observation every 6 to 10 minutes.  Submillimetre observations at
the SMTO used either of the single channel, double-sideband,
receivers (1.3mm or 0.87mm). Backends include the 250 kHz filterbanks,
which is used to display the results in this paper.  System
temperatures varied from 400 K to 1000 K depending on weather
conditions. Pointing/focus was done on a nearby planet and is
accurate to better than 2 arcseconds. Calibration was done with a hot
vane observation every 10 minutes.


\begin{figure*}
\centerline{\includegraphics[width=400pt]{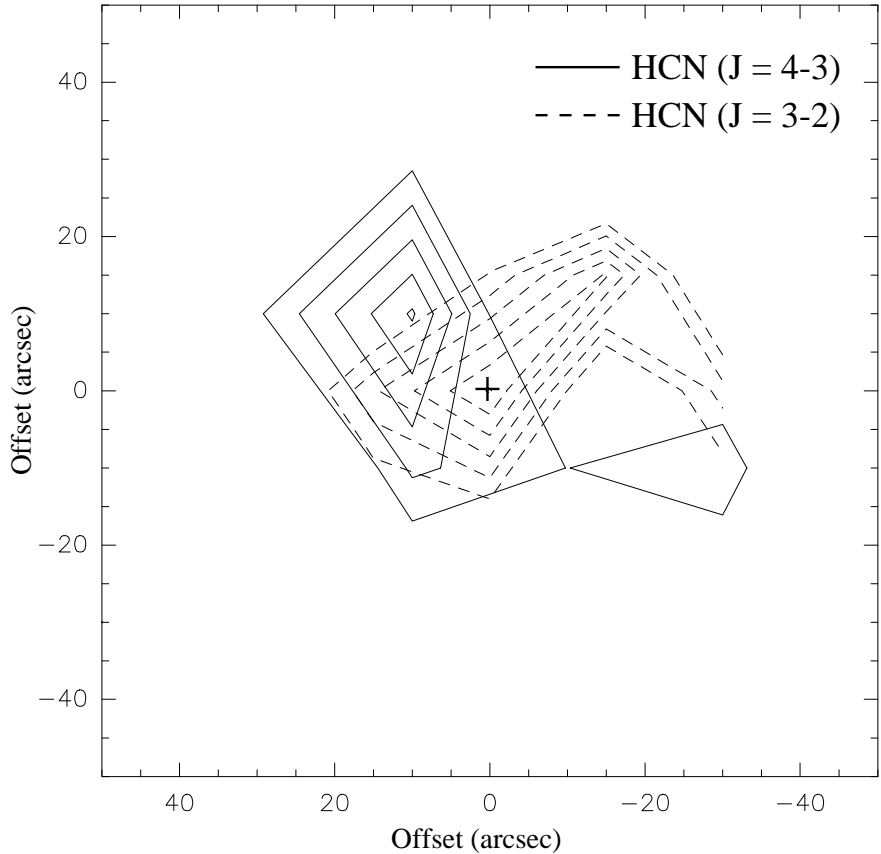}}
\caption{Maps of HCN emission in the coma of comet Ikeya-Zhang. Solid lines are
 HCN $J$=4-3 and broken lines are HCN $J$=3-2. The cross marks the
 nucleus position. For the $J$=4-3 map, the lowest contour is 3 times
 the rms value and the contours are in steps of one sigma
 (i.e. 3,4,5,6 and 7 sigma).  The peak flux is 3.6 K km/sec (7 sigma)
 and one sigma equals 0.51 K km/s. For the $J$=3-2 map, the lowest
 contour is 20 times the rms value and the contours are in steps of 3
 sigma each.  The peak flux is 0.9 K km/sec ( 35 sigma) and one sigma
 equals 0.026 K km/s. In both cases the sigma is the baseline rms per
 channel divided by the number of channels that are covered in the
 integration.}
\end{figure*}

Figure 2 shows a superposition of coma maps in the $J$=3-2 and $J$=4-3
 lines of HCN. The data was taken on different days (March 31 2002 for
 the HCN(4-3) map and April 10 2002 for the HCN(3-2) map). Both maps
 have been integrated between -2 and 2 km/sec of the line center,
 which represents the maximum line extent seen in either line.  The
 HCN(4-3) data are in the form of $T_A^*$ (which should be multiplied
 by 0.50 to convert to $T_R^*$) and the HCN(3-2) data are in the form
 of $T_R^*$. To convert the HCN(4-3) data to $T_R^*$, 1 sigma equals
 1.02 K km/s.  The beams were 21 arcseconds with 20 arcseconds spacing
 for $J$=4-3, and 24 arcseconds with 15 arcseconds spacing for
 $J$=3-2.

The offset between the centre of the map (0,0) and the HCN(4-3) peak
is real - about 10 arcseconds. We can rule out differences in
rotational emission for the observed $J$ levels since one would expect
that the HCN(3-2) emission would be more extended as one moves away
from the inner coma, where the gas density is falling.  The most
likely explanation is that when the HCN(4-3) observations were made an
extended source of HCN existed in the coma, whereas about 10 days
later this had dissipated and the major contributor to HCN emission,
as determined by the 3-2 line, was from molecules coming directly from
the nucleus.  This interpretation supports the idea that there are two
sources of HCN in cometary comae. One source is pristine interstellar
HCN from the nucleus, the other is photolytic decomposition of some
organic parent molecule. This parent may be the same as that which is
responsible for cometary HNC and hence, depending on the product
branching ratios for photolysis of this unknown molecule, the HNC/HCN
ratio in comets may actually reflect the nuclear ratio $\lbrack{\rm
organic~parent} \rbrack$/HCN.

We have also measured the HCN production rates at several epochs and
measured the HNC/HCN ratio at $r_h \sim 1$AU. These results will be
reported elsewhere, however we note here that the HNC/HCN ratio in
Ikeya-Zhang is $\sim 30$\%, and so is larger than measured in
Hale-Bopp at perhelion.


\section{Organic Interstellar Molecules in Comets} 

\begin{figure*}
\centerline{\includegraphics[width=350pt]{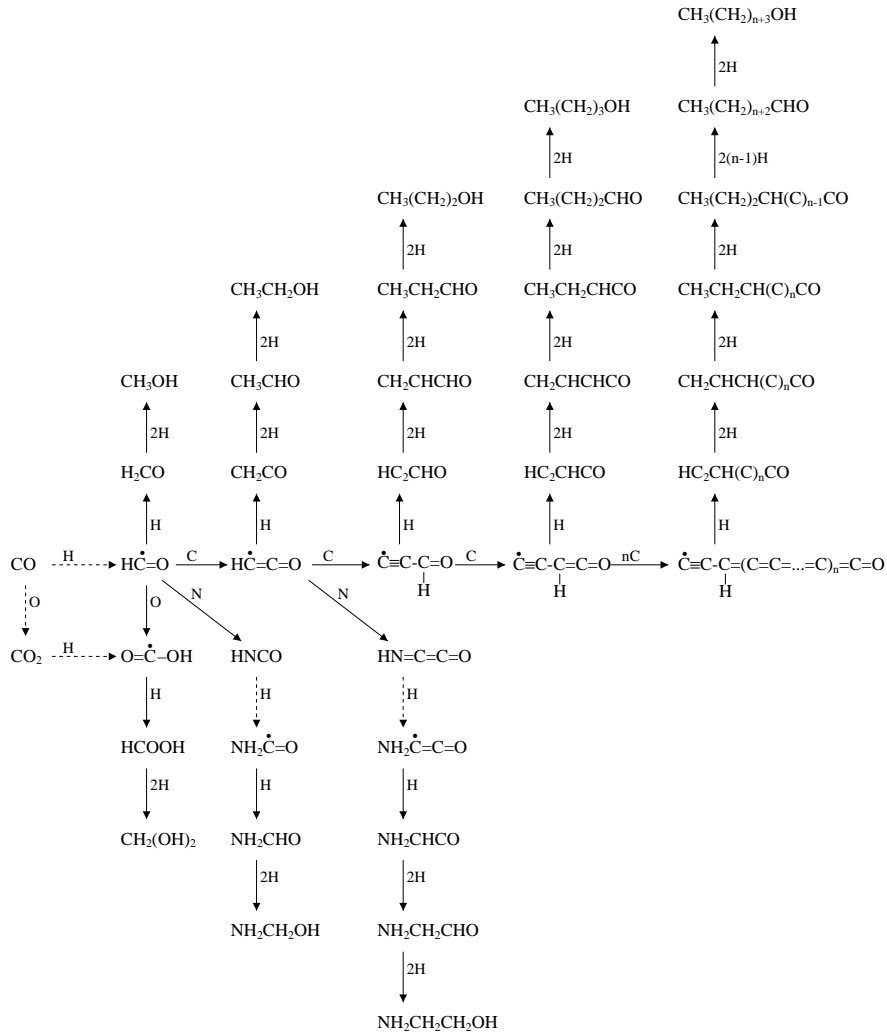}}
\caption{Interstellar grain surface chemistry (adapted from Charnley
2001).  Hydrogen atom addition to unsaturated molecules creates
reactive radicals and a rich organic chemistry seeded by carbon
monoxide ensues.  Broken arrows indicate reactions with activation
energy barriers; where 2H is shown, a barrier penetration reaction
followed by an exothermic addition is implicitly indicated.}
\end{figure*}


Table 1 shows that, although there are some discrepancies, there is a
strong correspondence between cometary molecules and those detected in
interstellar ice mantles or in warm gas containing these evaporated
ices - so-called `hot cores' (e.g.\ Charnley et al. 1992).  We can
strengthen this connection both by modelling coma organic chemistry
and by exploring the implications interstellar grain chemistry
theories have for cometary composition.

The brightness of comet Hale-Bopp allowed many organic molecules to be
positively detected (\BoMo\ \etal 2000). Many of the organic species
detected, such as HCOOCH$_3$, HC$_3$N, and CH$_3$CN, are also seen in
hot cores where they are thought to be daughter species formed by
gas-phase reactions (Rodgers \& Charnley 2001a). Thus, it is possible
these molecules were actually created in the coma of Hale-Bopp. Coma
modelling utilising the interstellar organic chemistry of these
molecules indicates that chemical reactions {\it cannot} form
sufficient quantities of HCOOH, HCOOCH$_3$, HC$_3$N and CH$_3$CN\@
(Rodgers \& Charnley 2001b). This result confirms that these species
originate in the nuclear ices and suggests that, prior to
incorporation into cometesimals, primordial interstellar grain mantles
ices were evaporated, perhaps within the protosolar nebula, underwent
a hot core-like phase of chemical processing, and then recondensed.


Figure 3 depicts chemistry driven by single atomic addition reactions
on grains, with the further constraint of {\it radical stability}
imposed on the intermediate organic radicals (Charnley 1997, 2001).
This scheme accounts for the presence in hot core precursor ices, and
presumably also cometary ices, of small organic molecules containing
the carbonyl group. The basic organic chemistry is driven primarily by
C and N additions. Further H additions lead to a saturated mantle rich
in alcohols although energy barriers for H additions to double and
triple bonds mean that complete saturation does not occur and some
aldehydes can persist. Acetaldehyde has been tentatively observed in
Hale-Bopp (Crovisier \etal 1999) and this theory suggests that,
depending of the reduction state of the cometary ices, ketene and/or
ethanol should also have been present.  Comets should also contain the
ring isomer of acetaldehyde, ethylene oxide (c-C$_2$H$_4$O).  Ethylene
oxide is formed by O atom addition to the C$_2$H$_3$ radical, which is
formed as an intermediate in the reduction of acetylene to ethene and
ethane; this sequence has been proposed as the origin of cometary
ethane (Mumma et al.\ 1996).  The HCOOH, HNCO and NH$_2$CHO observed
in both hot cores and comets are also accounted for by the
theory. Stronger support for a connection to cometary material comes
from PUMA mass spectrometry data of comet Halley.  Kissel \& Krueger
(1987) identified the mass peaks at $m/Z$= 44, 46 and 48 with
protonated HNCO (or HOCN), NH$_2$CHO and NH$_2$CH$_2$OH
(aminomethanol).


A'Hearn \etal (1995) demonstrated that short-period comets are much more
likely to be depleted in the carbon chains C$_2$ and C$_3$ than
long-period comets.  The current consensus is that Kuiper belt comets
most probably formed near their current location, whereas Oort cloud
comets originally formed somewhere in the giant planet region of the
protosolar nebula.  Understanding this differentiation may therefore
provide important cosmogonic information.  Recent work suggests that
C$_2$ and C$_3$ could be produced in the coma from C$_2$H$_2$,
C$_2$H$_6$ and C$_3$H$_4$ (Helbert et al. 2002, these proceedings).
Alternatively, the chemistry of Figure 3 lends support to the idea
that photolysis of nuclear propynal (HC$_2$CHO) could be the parent
of C$_3$ and some C$_2$ (Krasnopolsky 1991). Observational upper
limits on the nuclear abundance of propynal are not very restrictive
(Crovisier et al.\ 1993). One may speculate that C$_2$/C$_3$
differentiation amongst comet families, as well as other relative
molecular depletions (CO, CH$_3$OH, CH$_4$, C$_2$H$_2$, C$_2$H$_6$;
Mumma \etal 2000; Mumma \etal 2001; \BoMo\ \etal 2001), could be
traced back to differences on the environment in which the chemistry
of Figure 3 originally occurred. 


Ion-molecule chemistry at the temperatures of molecular clouds ($\sim
10-30$K leads to significant isotopic fractionation of D, $^{13}$C and
$^{15}$N in interstellar molecules (Tielens \& Hagen 1982; Langer et
al. 1984; Millar et al. 2000; Terzieva \& Herbst 2001). Hence,
measuring these ratios in cometary parent molecules can strengthen the
link with an interstellar provenance, and also offers the possibility
of extracting important information regarding the origin, pristinity
and formation temperature of the nuclear ices.  The only deuterated
molecules observed in comets are HDO and DCN\@. The observed
HDO/H$_2$O ratios lie in the range 5.7--6.6$\times10^{-4}$ (e.g.\
Meier \etal 1998a) and a DCN/HCN ratio of 0.002 was determined in
Hale-Bopp (Meier \etal 1998b).  However, rapid chemical reactions in
the coma could in principle alter D/H ratios, and this would
compromise any comparisons with interstellar chemistry. Comet
modelling of deuterium chemistry in the coma has demonstrated that, as
in hot cores, post-evaporation chemistry does not significantly alter
the initial D/H ratios (Rodgers \& Charnley 2002a). Therefore, for
parent species we can use the observed coma ratios to infer the
nuclear D/H ratio, and for daughter species we can use these ratios to
constrain the nature of the parent and the chemical mechanism by which
the species are formed. In particular, if the DNC/HNC ratio can be
determined in a comet, we may be able to ascertain the origin of
cometary HNC, which is currently not well understood (see above).

\section{Conclusions}

Modelling of coma chemistry can constrain nuclear composition and
isotopic fractionation, as well as the identification of the most
important formation processes for individual molecules. We have
highlighted this approach in studies of the important HNC/HCN
ratio. More measurements of this ratio as a function of heliocentric
distance are needed, as are maps of the individual
molecules. Reactions involving suprathermal hydrogen atoms is an
efficient pathway for HNC production in some comets. Fast H atoms may
play an important general role in inner coma chemistry, particularly
so in the sulphur chemistry where the barrier for H abstraction from
H$_2$S is the lowest amongst the most abundant parent molecules.

Existing measurements of the $^{15}$N/$^{14}$N ratio in CN and HCN
(Crovisier 1999) show them to be consistent with bulk Solar System
values.  The measurement of an enhanced $\rm ^{15} NH_3/ ^{14}NH_3 $
ratio in a comet would provide positive evidence for the interstellar
fractionation scenario proposed by Charnley \& Rodgers (2002) which
also gives a possible explanation of the widespread depletion of N$_2$
found in comets.

The detection of even more deuterated species would greatly constrain
the connection with interstellar molecules. In particular, recent
observations of low-mass protostellar cores yield very large
HDCO/H$_2$CO and D$_2$CO/H$_2$CO ratios (Loinard \etal 2001) - measuring
formaldehyde deuteration should be an important goal for the
future. Determination of the DNC/HNC ratio would constrain some coma
chemistry scenarios (Rodgers \& Charnley 2002a).

Anomalous water emission has recently been discovered in the inner
envelope of the carbon-rich AGB star IRC+10216 (Melnick et
al. 2001). This has been attributed to the sublimation of a population
of comets, residing at a radius corresponding to that of the Kuiper
Belt in the Solar System.  The discovery of HDO molecules in IRC+10216
would be the definitive test of this scenario, as stars on the AGB
have consumed their deuterium (Rodgers \& Charnley 2002b).  As HDO
molecules would also be present during the O-rich phase of evolution,
it would also be an excellent general probe of extrasolar cometary
material at the final stages of stellar evolution.

Finally, cometary ortho:para spin ratios have been measured for H$_2$O
in a number of comets (Mumma \etal 1993; Irvine \etal 2000) and for
H$_2$CO, CH$_4$, NH$_3$(NH$_2$) and CH$_3$OH in Hale-Bopp (Crovisier
1999; Kawakita et al.\ 2001). The observed ratios imply formation
temperatures for cometary molecules of typically 25--35~K; this range
has important cosmogonic implications.  However, it is not certain
that the observed values accurately reflect the ratios in the
nucleus. The presence of chemically significant abundances of H$^+$
and protonated ions in the inner coma leads to the possibility that
rapid ion-molecule spin-exchange (Dalgarno \etal 1973) and proton
transfer reactions (Kahane \etal 1984) could alter these ratios from
those originally present in the nucleus. Detailed models of the coma
chemistry are necessary in order to ensure that coma observed
ortho:para ratios accurately reflect the nuclear values.


\acknowledgements This work was supported by NASA's Exobiology and
 Planetary Atmospheres Programs, through NASA Ames Interchange
 NCC2-1162, and by the Netherlands Research School for Astronomy
 (NOVA).




\def\refindent{\par\noindent\parskip=0pt\hangindent=3pc\hangafter=1 }
\def\apj#1#2#3{\refindent#1,  {\it Ap.J., }{\bf#2}, #3.}
\def\apjlett#1#2#3{\refindent#1,  {\it Ap.J.(Letters), }{\bf#2}, #3.}
\def\apjsupp#1#2#3{\refindent#1,  {\it Ap.J.Suppl., }{\bf#2}, #3.}
\def\apsp#1#2#3{\refindent#1,  {\it Ap.Sp.Sci.,}{\bf#2},#3.}
\def\mn#1#2#3{\refindent#1,  {\it M.N.R.A.S., }{\bf#2}, #3.}

\def\mns#1#2#3{\refindent#1,  {\it M.N.R.A.S. }{\bf#2}, #3.}
\def\aj#1#2#3{\refindent#1,  {\it A.J., }{\bf#2}, #3.}
\def\aa#1#2#3{\refindent#1,  {\it Astr.Ap., }{\bf#2}, #3.}
\def\Nature#1#2#3{\refindent#1,  {\it Nature, }{\bf#2}, #3.}
\def\Icarus#1#2#3{\refindent#1,  {\it Icarus, }{\bf#2}, #3.}
\def\refpaper#1#2#3#4{\refindent#1,  {\it #2 }{\bf#3}, #4.}

\def\capri#1#2#3#4#5{\refindent#1, {\rm #2 }, $\underline{ #3 },~${\bf#4}, #5}
\def\cbook#1#2{\refindent#1, {\it #2} }
\def\refbook#1{\refindent#1}
\def\science#1#2#3{\refindent#1, {\it Science, }{\bf#2}, #3.}

\newcommand{\coref}[6]{\bibitem[]{} #1, #6, {\it #3}, {\bf #4}, #5}

\newcommand{\cobook}[7]{\item #1, #2, in {\underline{#3}}, eds.\ #4,
#5, p.~#6, #7}


\end{article}

\begin{thebibliography}{}

\coref{A'Hearn M.F., Millis R.L., Schleicher D.G., Osip D.J., Birch P.V.}
{The Ensemble properties of Comets: Results from Narrowband Photometry
of 85 Comets, 1976--1992}{Icarus}{118}{223-270}{1995}

\coref{Bernstein M.P., Sandford S.A., Allamandola L.J., Chang S., Scharberg M.A.}
{TITLE}{ApJ}{454}{327}{1995}

\coref{Blake G.A., Qi C., Hogerheijde M.R., Gurwell M.A., Muhleman D.O.}
{Sublimation from Icy Jets as a Probe of the Interstellar Volatile
Content of Comets}{Nature}{398}{213-216}{1999}

\coref{\BoMo\ D. et al.}
{New Molecules Found in Comet C/1995 O1 (Hale-Bopp) -- Investigating
the Link Between Cometary and Interstellar Material}{A\&A}{353}{1101-1114}{2000}

\bibitem{}\BoMo\ D. et al., 2001, {\it Science}, {\bf 292}, 1339-1443

\bibitem{} Charnley S.B. 1997, 
in C.B. Cosmovici, S. Bowyer \& D. Werthimer (eds.), {\it Astronomical and Biochemical Origins and the
Search for Life in the Universe}, Editrice Compositori, Bologna, p.~89

\bibitem{} Charnley S.B. 2001, 
in F. Giovannelli (ed), {\it The Bridge Between the Big Bang and Biology},
Consiglio Nazionale delle Ricerche, Italy, p.~139

\coref{Charnley S.B. and Rodgers S.D.}{15N and the end of ISM chem}
{ApJ (Letters)}{569}{L133-137}{2002}

\coref{Charnley S.B., Tielens A.G.G.M., Millar T.J.}
{On the Molecular Complexity of the Hot Cores in Orion A: Grain
Surface Chemistry as ``The Last Refuge of the Scoundrel''}{ApJ}{399}{L71-74}{1992}

\bibitem{}Charnley S.B., Rodgers S.D., Kuan Y-J., Huang H-C., 2002,
  {\it Adv.\ Space Res.}, in press

\bibitem{}Crovisier J., 1999, in J.M. Greenberg \& A. Li (eds.), {\it
Formation and Evolution of Solids in Space}, Kluwer, Dordrecht, pp.~389-426

\coref{Crovisier J., \BoMo\ D., Colom P., Despois D., Paubert G.}{TITLE!!!}
{A\&A}{269}{527-540}{1993}

\coref{Crovisier J. et al.}{TITLE!!!}{AAS DPS}{\#31}{\#32.02}{1999}

\bibitem{}Dalgarno A., Black J.H., Weisheit J.C., 1973, {\it Astrophys.\
  Lett.}, {\bf 14}, 77-79

\coref{Ehrenfreund P. and Charnley S.B.}{Organic Molecules in the
Interstellar Medium, Comets and Meteorites: A Voyage from Dark Clouds
to the Early Earth}{ARA\&A}{38}{427-483}{2000}

\bibitem{}Huebner W.F., 1985, in Diercksen et al.\ (eds.), {\it Molecular Astrophysics},
 Reidel, Dordrecht, pp.~311-330

\bibitem{}Hunter E.P. and Lias S.G., 1998, in 
  W.G. Mallard \& P.J. Linstrom (eds.), {\it 
NIST Chemistry  WebBook, NIST Standard Reference Database No.\ 69},
NIST, Gaithersburg MD, (http://webbook.nist.gov)

\coref{Irvine W.M. et al.}
{Spectroscopic Evidence for Interstellar Ices in Comet Hyakutake}
{Nature}{383}{418-421}{1996}

\coref{Irvine W.M. et al.}
{Chemical Processing in the Coma as the Source of Cometary HNC}
{Nature}{393}{547-550}{1998}

\bibitem{} Irvine W.M., Schloerb F.P., Crovisier J., Fegley B., Mumma
M.J., 2000, in V. Mannings, A.P. Boss \& S.S Russell (eds.), 
{\it Protostars and Planets IV}, University of Arizona Press, Tucson, pp.~1159-1200

\coref{Kahane C., Lucas R., Frerking M.A., Langer W.D., Encrenaz P.}{TITLE}
{A\&A}{137}{211-222}{1984}

\coref{Kawakita H. et al.}{TITLE!!!!!}{Science}{294}{1089-1091}{2001}

\coref{Kissel J., and Krueger F.R.}{TITLE!}{Nature}{326}{755-760}{1987}

\bibitem{}\Ko~A. et al., 1987, {\it J. Geophys.\ Res.}, {\bf 92}, 7331-7340

\coref{Krasnopolsky V.A.}{TITLE!!!!!!}{A\&A}{245}{310-315}{1991}

\coref{Langer W.D., Graedel T.E., Frerking M.A., Armentrout P.B.}
{Carbon and Oxygen Isotope Fractionation in Dense Interstellar Clouds}
{ApJ}{277}{581-604}{1984}

\coref{Loinard L., Castets A., Ceccarelli C., Caux E., Tielens
  A.G.G.M.}{Doubly deuterated molecular species in protostellar
  environments.}{ApJ}{552}{L163-166}{2001}

\coref{Meier R. et al.}
{A Determination of the HDO/H$_2$O Ratio in Comet C/1995 O1
(Hale-Bopp)}{Science}{279}{842-846}{1998a}

\coref{Meier R. et al.}
{Deuterium in Comet C/1995 O1 (Hale-Bopp): Detection of DCN}
{Science}{279}{1707-1710}{1998b}

\coref{Melnick G.J., Neufeld D.A., Ford K.E.S., Hollenbach D.J.,
  Ashby M.L.N.}{Discovery of water vapor around IRC10216 as evidence for comets orbiting another star.} 
{Nature}{412}{160-163}{2001} 

\coref{Millar T.J., Roberts H., Markwick A.J., Charnley S.B.}{The Role
of H$_2$D$^+$ in the Deuteration of Interstellar Molecules}{Phil.\
Trans.\ R. Soc.\ Lond. A.}{358}{2535-2547}{2000}

\bibitem{}Mumma M.J., Weissman P.R., Stern S.A., 1993, in E.H. Levy \& J. Lunine (eds),
{\it Protostars and Planets III}, University of Arizona Press, Tucson, pp.~1177-1252

\bibitem{}Mumma M.J., DiSanti M.A., Dello Russo N., Fomenkova M.,
  Magee-Sauer K., Kaminski C.D., Xie D.X., 1996, {\it Science},
  {\bf 272}, 1310-1314

\coref{Mumma M.J., DiSanti M.A., Dello Russo N., Magee-Sauer K.,
Rettig T.W.}{Detection of CO and Ethane in Comet 21P/Giacobini-Zinner:
Evidence for Variable Chemistry in the Outer Solar Nebula}{ApJ}{531}{L155-159}{2000}

\coref{Mumma M.J. et al.}{TITLE!!!!!!!!!}{Science}{292}{1334-1339}{2001}

\coref{Rettig T.W., Tegler S.C., Pasto D.J., Mumma M.J.}{Comet Outbursts
and Polymers of HCN}{ApJ}{398}{293-298}{1992}

\coref{Rodgers S.D. and Charnley S.B.}{HNC and HCN in Comets}
{ApJ}{501}{L227-230}{1998}

\coref{Rodgers S.D. and Charnley S.B.}{Chemical Differentiation in
Regions of Massive Star Formation}{ApJ}{546}{324-329}{2001a}

\coref{Rodgers S.D. and Charnley S.B.} {Organic Synthesis in the Coma
of Comet Hale-Bopp?}{MNRAS}{320}{L61-64}{2001b}

\coref{Rodgers S.D. and Charnley S.B.}{On the Origin of HNC in Comet
Lee}{MNRAS}{323}{84-92}{2001c}

\coref{Rodgers S.D. and Charnley S.B.}{A Model of the Chemistry in
Cometary Comae: Deuterated Molecules}{MNRAS}{330}{660-674}{2002a}

\bibitem{}Rodgers S.D. and Charnley S.B. 2002b, 
{\it Planet.\ Space Sci.}, In press

\coref{Schmidt H.U., Wegmann R., Huebner W.F., Boice D.C.}{!!!!}
 {Comp.\ Phys.\ Comm.}{49}{17}{1988}

\coref{Talbi D., Ellinger Y., Herbst E.}{On the HNC/HCN Abundance
Ratio: a Theoretical Study of the H $+$ CNH $\leftrightarrow$ HCN $+$
H Exchange Reactions}{A\&A}{314}{688-692}{1996}

\coref{Terzieva R. and Herbst E.}{The Possibility of Nitrogen Isotopic
Fractionation in Interstellar Clouds}{MNRAS}{317}{563-568}{2000}

\coref{Tielens A.G.G.M. and Hagen W.}
{Model Calculations of the Molecular Composition of
Interstellar Grain Mantles }{A\&A}{114}{245-260}{1982}

\coref{Veal J.M. et al.}{An interferometric study of HCN in comet Hale-Bopp.}
{AJ}{119}{1498-1511}{2000}


























\end{thebibliography}
\end{document}